\documentclass[twocolumn,tighten]{aastex63}
\usepackage{amsmath}

\shorttitle{External Field Effect in disk galaxies}
\shortauthors{Chae and Milgrom}

\begin{document}
\title{Numerical Solutions of the External Field Effect on the Radial Acceleration in Disk Galaxies}

\correspondingauthor{Kyu-Hyun Chae}

\author{Kyu-Hyun Chae}
\affil{Department of Physics and Astronomy, Sejong University, 209 Neungdong-ro Gwangjin-gu, Seoul 05006, Republic of Korea}
\email{KC: chae@sejong.ac.kr, kyuhyunchae@gmail.com}

\author{Mordehai Milgrom}
\affil{Department of Particle Physics and Astrophysics, Weizmann Institute of Science, Rehovot 7610001, Israel}
\email{MM: moti.milgrom@weizmann.ac.il}

\begin{abstract}
In MOND (modified Newtonian dynamics)-based theories the strong equivalence principle is generically broken in an idiosyncratic manner, manifested in the action of an ``external field effect (EFE)''. The internal dynamics in a self-gravitating system is affected even by a constant external field. In disk galaxies the EFE can induce warps and modify the rotational speeds. Due to the non-linearity of MOND, it is difficult to derive analytic expressions of this important effect in a disk. Here we study numerically the EFE in two non-relativistic Lagrangian theories of MOND: the `Aquadratic-Lagrangian' theory (AQUAL) and `Quasilinear MOND' (QUMOND). For AQUAL we consider only the axisymmetric field configurations with the external field along the disk axis, or a spherical galaxy with test-particle orbits inclined to the external field. For the more manageable QUMOND we calculate also the three-dimensional field configurations, with the external field inclined to the disk axis. We investigate particularly to what degree an external field modifies the quasi-flat part of rotation curves. While our QUMOND results agree well with published numerical results in QUMOND, we find that AQUAL predicts weaker EFE than published AQUAL results. However, AQUAL still predicts stronger EFE than QUMOND, which demonstrates current theoretical uncertainties. We also illustrate how the MOND prediction on the rising part of the rotation curve, in the inner parts, depends largely on disk thickness but only weakly on a plausible external field for a fixed galaxy model. Finally, we summarize our results for the outer parts as an improved, approximate analytic expression.    
\end{abstract}


\section{Introduction} \label{sec:intro}
Modified Newtonian dynamics (MOND) is an alternative to the dark matter (DM) hypothesis in the Universe. General relativity and its Newtonian limit require invoking large quantities of dark matter to account for the observed dynamics of galactic systems and of the Universe at large. MOND modifies general relativity and Newtonian dynamics at low accelerations, at or below $\sim 10^{-10}{\rm m~s^{-2}}$, in a way that eliminates the need to invoke non-baryonic dark matter. MOND is reviewed, e.g., in \cite{FM2012}, \cite{Mil2014a}(updated continually), \cite{merritt2020}, and \cite{BZ2021}.

General relativity is conjectured to be the only full-fledged relativistic theory of gravity that obeys the strong equivalent principle (SEP). Theories that depart from general relativity can break the SEP in different ways. In particular, it was recognized right from the advent of MOND that it breaks the SEP in a specific way, manifested by the action of an ``external field effect'' (EFE) \citep{Mil1983}. This important difference may provide a way to distinguish between MOND and dark matter as the explanation of the mass discrepancies in galactic systems.

In the astrophysical context, the effect may concern various small subsystems that are falling freely in the gravitational field of a larger mother system, such as binary stars, star clusters, and satellite galaxies in the field of a galaxy, or a galaxy falling in the field of a galaxy cluster or of the general surrounding large-scale structure. 

According to the SEP, the internal dynamics of such a subsystem is oblivious to the presence of a constant gravitational acceleration field of the mother system (which can only make itself felt by tidal effects if it varies across the subsystem).

In the MOND paradigm, however, even a constant external field may affect the internal dynamics (see below). The presence of this effect has been explicitly demonstrated and discussed in detail in specific MOND formulations: the initial `algebraic' formulation \citep{Mil1983}, the ``Aquadratic-Lagrangian'' (AQUAL) formulation \citep{BM1984}, the ``Quasilinear'' formulation (QUMOND) \citep{Mil2010}, and in ``modified-inertia'' formulations (e.g., \citealt{Mil2011}). In fact, it was shown that although not totally unavoidable, the EFE is generic to MOND  \citep{Mil2014b}.

There have been observational indications of the action of the EFE in various galactic systems, e.g., dwarf satellites of the Andromeda galaxy, which are subject to the field of Andromeda \citep{mm13}, disk galaxies from the Spitzer Photometry and Accurate Rotation Curves (SPARC; \citealt{Lel2016}) database in the fields of external mass concentrations, and ultra-diffuse galaxies (UDGs) in the field of the Coma cluster. \cite{Chae2020,Chae2021} have used a simple approximate model of the EFE \citep{FM2012}, analyzing the rotation curves (RCs) of more than 150 SPARC galaxies and inferred the values of the external fields in which they fall. They have pointed out that these deduced fields are correlated with the large-scale cosmic environments of galaxies, supporting the EFE interpretation of the SPARC data. On the other hand, \cite{Freund2021} have not found indications of EFE from an analyses of 11 UDGs in the Coma cluster (a fact that could, however, be due to their being heated tidally in the cluster), while study of tidally-affected dwarfs in the Fornax cluster found, to the contrary, strong indication that they are subject to the EFE of the cluster (E. Asencio, private communication).

Testing the EFE using RCs or stellar velocity dispersions of galaxies requires both accurate data and realistic modeling. Published analyses of RCs (e.g.\ \citealt{Haghi2016,Hees2016,Chae2020,Chae2021}) were based on an approximate analytical formula from \citep{FM2012}, or on numerical simulation results for {\it spherical} galaxies \citep{Haghi2019}. These models are first-order approximations in that the one-dimensional model of \cite{FM2012} is based on the assumption that the external field is always aligned with the internal radial acceleration, and the simulation results by \cite{Haghi2019} are for velocity dispersions in spherical systems. Clearly, more realistic models than these are needed. 

Some more recent papers presented numerical results of the MOND EFE in disks \citep{Zon2021,Oria2021}. These studies used the QUMOND formalism and considered the case that the external field is parallel to the rotation axis. 

Here we extend these studies in QUMOND to the more realistic non-axisymmetric (3-dimensional) gravitational field configurations  where the external field is not aligned with the rotation axis. We also carry out numerical computations in the numerically less tractable AQUAL formalism to compare the predictions of the two versions of MOND. Some such numerical studies in AQUAL were described in \cite{BM2000}. In the present study of AQUAL,  we consider disks of various thicknesses and scale lengths in an external field parallel to the rotation axis. The effect of a tilted external field is studied only using orbits tilted with respect to the external field in a spherical system. Our results will provide possible ranges of the EFE on the galactic RCs based on currently available theories. We also illustrate what MOND predicts about internal accelerations within the disk radii with and without EFE, in particular for galaxies, such as dwarf and low surface brightness (LSB) galaxies, whose observed internal accelerations are very weak ($< \sim 10^{-10.5}$~m~s$^{-2}$) at all radii. Our results will be useful for refining studies of the EFE in rotationally-supported galaxies \citep{Chae2020,Chae2021}.

\section{Methodology} \label{sec:meth}
We shall be studying the EFE on spherical and on disk-like matter distributions, in two ``modified-gravity'', nonrelativistic formulations of MOND: AQUAL  \citep{BM1984}, and QUMOND \citep{Mil2010}. The gravitational degree of freedom in both theories is the nonrelativistic MOND potential, $\Phi$. In both theories, $\Phi$ is determined from the mass distribution (of baryons), $\rho$, by field equations that are generalizations of the Newtonian Poisson equation. 
In AQUAL, the field equation is 
\begin{equation}
  \mathbf{\nabla} \cdot \left[ \mu\left(|\mathbf{\nabla}\Phi|/a_0\right)\mathbf{\nabla}\Phi \right] = 4\pi G\rho,
 \label{eq:aqual}
\end{equation}
and in QUMOND the equation is 
\begin{equation}
  \mathbf{\Delta}\Phi=\mathbf{\nabla}\cdot \left[ \nu\left( |\mathbf{\nabla}\Phi^\ast_{\rm N}|/a_0 \right) \mathbf{\nabla}\Phi^\ast_{\rm N} \right].
    \label{eq:qumond}
\end{equation}
Here, $\Phi^\ast_{\rm N}$ is the total Newtonian potential sourced by $\rho$ including the external (mother) system if present (the symbol $\Phi_{\rm N}$ is reserved to denote the internal Newtonian potential of the subsystem) and through which $\rho$ enters the QUMOND equation, $G$ is Newton's constant, and $a_0\approx 1.2\times 10^{-10}{\rm m~s^{-2}}$ \citep{MLS2016} is the MOND acceleration constant; we shall work hereafter in units where $a_0=G=1$. The functions $\mu(x)$ and $\nu(y)$ are, respectively, the ``interpolating functions (IFs)'' for the two theories, with the asymptotic behavior at small and large argument values dictated by the basic tenets of MOND (see the above mentioned MOND reviews): $\mu(x\gg 1)\approx 1$, $\mu(x\ll 1)\approx x$, $\nu(y\gg 1)\approx 1$, $\nu(y\ll 1)\approx y^{-1/2}$.

Both field equations are derived from an action principle obeying the standard symmetries (space-time translations and rotations); so the theories enjoy the standard conservation laws (energy, momentum, and angular momentum). Both theories are gotten as the nonrelativistic limit of a relativistic formulation of MOND that predicts the correct (as observed) gravitational lensing.

Despite the different appearances, the two theories are rather similar in many regards; and, by and large, make similar predictions on most observables studied to date, although they do differ in quantitative detail. In fact, the two theories are related in the sense explained in \cite{Mil2012a}.

The similarity is established if we use in the two theories IFs that are related to each other as follows: For a given $\mu(x)$, invert the relation $x\mu(x)=y$, then define
\begin{equation}
 \nu(y)=x(y)/y, ~~~ \text{or} ~~~\mu(x)\nu(y)=1.
    \label{eq:munu}
\end{equation}
With this matching choice, the two theories are, for example, completely equivalent in configurations of high symmetry (spherical, cylindrical, and plane symmetric).

While AQUAL is a fully nonlinear theory, QUMOND requires solving only the linear Poisson equation twice: once to calculate $\mathbf{\nabla}\Phi_{\rm N}$ from $\rho$, and another to calculate $\mathbf{\nabla}\Phi$ from Equation~(\ref{eq:qumond}). QUMOND is thus much more amenable to numerical calculations. 

The EFE has been derived analytically for the asymptotic regime, far from a bounded mass in AQUAL by \cite{BM1984}, and in QUMOND by \cite{Mil2010} and by \cite{BZ2015}. Based on the asymptotic potentials given in these papers, \cite{Chae2021} summarize the azimuthally-averaged radial acceleration in a orbital plane whose axis is tilted by an angle of $\theta$ from a constant external field. From the Appendix~A of \cite{Chae2021} the AQUAL theory has the asymptotic radial acceleration far away from a bounded mass distribution:
\begin{equation}
  g|_{\rm A,asym} = g_{\rm N}\frac{1}{\mu(x)}  \left. \left( 1+\hat\mu(x) - \hat\mu(x) \frac{\sin^2\theta}{2} \right)^{-1/2} \right|_{x=e},
  \label{eq:pointa}
\end{equation}
where $g_{\rm N}=M/r^2$ is the asymptotic internal Newtonian acceleration, $e$ is the MOND external field in units of $a_0$, and $\hat{\mu}(x) \equiv d\ln\mu(x)/d\ln x$. The QUMOND theory has 
\begin{equation}
  g|_{\rm Q,asym} = g_{\rm N} \nu(y)  \left. \left(1+\frac{\hat{\nu}(y)}{2} - \hat{\nu}(y) \frac{\sin^2\theta}{4}\right)\right|_{y=e_{\rm N}},
  \label{eq:pointq}
\end{equation}
where $e_{\rm N}$ is the Newtonian external field in units of $a_0$ and $\hat{\nu}(y)\equiv d \ln\nu(y)/d\ln y$.  If $\mu$ and $\nu$ are related as in Equation~(\ref{eq:munu}), then $\hat\nu=-\hat\mu/(1+\hat\mu)$. These analytic expressions for the asymptotic limits provide important checks of numerical solutions.

Another often-used formulation of MOND -- to be referred to as the `algebraic' formulation -- relates the MOND and Newtonian accelerations at a given position algebraically, as 
\begin{equation}
 \mu(g/a_0)g=g_{\rm N},~~ {\rm or ~equivalently}~~ g=\nu(g_{\rm N}/a_0)g_{\rm N},
    \label{eq:algeb}
\end{equation}
with $\nu$ and $\mu$ related as in Equation~(\ref{eq:munu}). This was the original formulation of MOND, and it captures several salient properties and predictions of MOND (including an EFE). While it cannot have a general validity, this relation was shown to be the exact relation for rotation curve (but not for other aspects of the dynamics) in so-called modified-inertia theories of MOND \cite{Milgrom1994}. It is routinely used to predict MOND rotation curves and is in the basis of the zero-scatter MDAR (or RAR). It does predict somewhat different RCs than either AQUAL or QUMOND (themselves predicting somewhat mutually different RCs), as shown e.g. in \cite{BM1995,Mil2012b,LB2021}, and as we find here (see below). For one-dimensional systems (e.g., spherical) AQUAL and QUMOND coincide with the algebraic relation.

For the sake of concreteness and of simplicity we base our galaxy models on an axisymmetric Miyamoto-Nagai (MN; \citealt{MN1975}) mass distribution (or a combination of MN models) whose Newtonian potential is given in the cylindrical coordinates by 
\begin{equation}
    \Phi_{\rm N}(R,z) = - [R^2+(a+\sqrt{z^2+b^2})^2]^{-1/2},
    \label{eq:PhiN}
\end{equation}
where $a$ and $b$ are the MN parameters. All lengths are given in units of the MOND radius $r_{\rm M}\equiv \sqrt{GM/a_0} =\sqrt{M}$ in our units, where $M$ is the total mass of the disk. All calculated accelerations are then in units of $a_0$. The corresponding MN axisymmetric density $\rho(R,z)$ is gotten from the Poisson equation:
\begin{equation}
    \rho(R,z)=\frac{Mb^2}{4\pi}\frac{aR^2+(a+3\sqrt{z^2+b^2})(a+\sqrt{z^2+b^2})^2}{[R^2+(a+\sqrt{z^2+b^2})^2]^{5/2}(z^2+b^2)^{3/2}},
    \label{eq:rho}
\end{equation}
whose density contours can be found, e.g., in section~2.3.1 of \cite{BT2008}. The MN distribution can be used to approximate various models of mass distribution by considering different parameter values. For example, it becomes a Plummer sphere when $a=0$. In the limit $b\rightarrow 0$ the MN distribution goes to a thin Kuzmin disk \citep{BT2008}. The MN model also allows an analytic expression for the QUMOND effective density (Appendix~\ref{sec:math}) which is useful for our QUMOND algorithm of three-dimensional integration.

\begin{figure}
  \centering
  \includegraphics[width=1.0\linewidth]{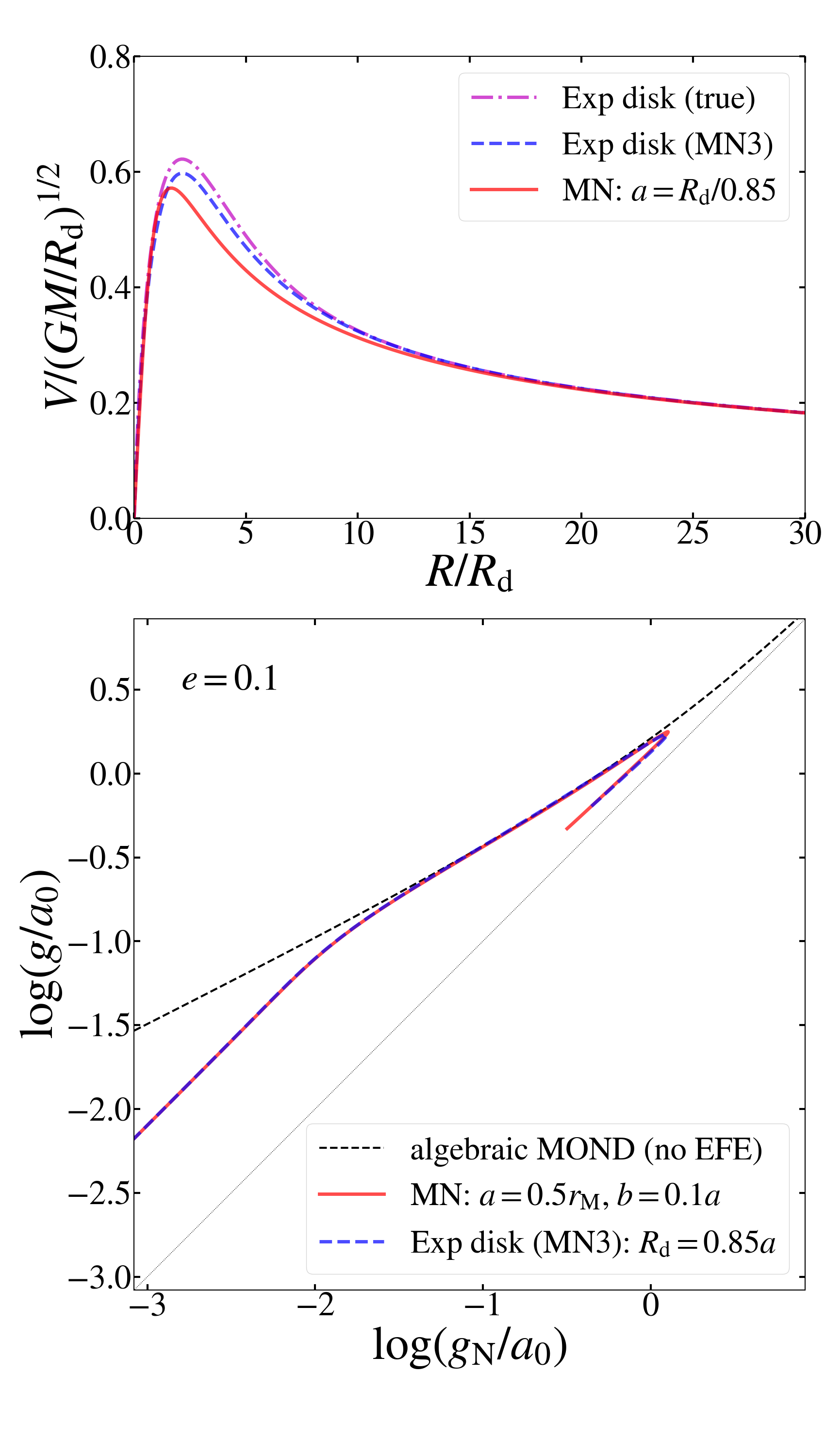}
    \vspace{-1.truecm}  
    \caption{\small 
    \textbf{MN versus exponential disk.}
 Top panel: Newtonian rotation curves of the exponential disk (magenta dash-dot), an exponential disk approximated by a combination of three MN models (blue dashed) using the prescription of \cite{Smith2015}, and the MN disk with zero thickness (i.e.\ Kuzmin disk, red solid) are compared. Here the MN disk scale length is set to be $a=R_{\rm d}/0.85$. The approximate disk model of \cite{Smith2015} is adjusted to have the correct total mass. The MN model with $a=R_{\rm d}/0.85$ has 8\% lower maximum rotation speed at 19\% smaller radius compared with the exponential disk. The approximate exponential disk agrees with the true exponential disk within 5\%.  Bottom panel: AQUAL-based numerical RARs for two of the above disks under an external field parallel to the rotation axis are nearly indistinguishable for the Kuzmin disk and the approximate exponential disk for $R_{\rm d}=0.85 a$. 
  }
   \label{MN_expdisk}
\end{figure}

It can also be used to approximate the more realistic exponential disk of any thickness. The parameter $a$ is similar to the disk scale length ($R_{\rm d}$) in that the maximum Newtonian rotation speeds occur at $\sim 1.5a$ and $\sim 2R_{\rm d}$. The ratio $b/a$ controls the disk thickness. Of course, real disk galaxies cannot be accurately described by an MN model (or even an exponential disk) alone. However, our goal here is to study how the EFE modifies the RC in a disk plane, rather than fitting the RC with a model. We have checked for the case of external field parallel to the disk axis that our numerical results for the MN model are nearly indistinguishable from those based on an exponential disk model approximated by a combination of three MN models using the prescription of \cite{Smith2015}. Figure~\ref{MN_expdisk} explicitly shows that the AQUAL numerical RAR of an MN model is nearly identical to that of an approximate exponential disk model with $R_{\rm d}=0.85 a$, i.e., when two models are slightly adjusted to cover the same acceleration range.  

In deriving various EFE simulation results in both AQUAL and QUMOND and comparing them each other, we will use MN models. However, when we consider specific examples of galaxies in AQUAL, we will use exponential disk models approximated by a combination of three MN models. We consider the case that the galaxy is freely falling in a constant external acceleration field. The $z$ axis is taken along the rotation axis of the disk, and the external field makes an angle of $\theta$ with it. 

\subsection{QUMOND algorithm}
Taking $\Phi_{\rm N}$ to be the Newtonian potential due only to the galaxy, the right-hand side of the QUMOND field equation (\ref{eq:qumond}) can be written as $4\pi \hat\rho$, where $\hat\rho$ is an effective density source
\begin{equation}
    \hat{\rho}= \frac{1}{4\pi} \mathbf{\nabla}\cdot \left[ \nu\left( |\mathbf{\nabla}\Phi_{\rm N}-\mathbf{g}_{\rm N,ext}| \right) (\mathbf{\nabla}\Phi_{\rm N}-\mathbf{g}_{\rm N,ext}) \right],
    \label{eq:rhohat}
\end{equation}
where $\mathbf{g}_{\rm N,ext}=e_{\rm N} (\sin\theta \hat{y} + \cos\theta \hat{z})\equiv e_{{\rm N}y}\hat{y} + e_{{\rm N}z}\hat{z}$ is the Newtonian external field, which is parallel to the $yz$-plane. Its magnitude is $e_{\rm N}$, and its components along the $y$ and $z$ axes are $e_{{\rm N}y}$ and $e_{{\rm N}z}$ respectively. For $\nu(y)$ we take the ``simple'' form $\nu(y)=1/2+\sqrt{1/4+1/y}$ \citep{FB2005}. (While the behavior of this $\nu$ at very high accelerations is not consistent with solar-system constraints, it is a very good approximation and describes rotation curves well in the low-to-intermediate-acceleration regime considered here.)

The difference $\rho_p\equiv\hat\rho-\rho$ is termed the ``phantom density'' because to get the MOND potential $\Phi$ in a Newtonian framework one would need to invoke $\rho_p$ as the ``dark matter'' density distribution \citep{Mil1986b}. This is a useful concept especially for comparing MOND predictions directly with dark matter inferences. For example, \cite{Oria2021} have recently derived the phantom density distribution in the local volume using QUMOND.

 For the MN Newtonian potential of Equation~(\ref{eq:PhiN}), $\hat\rho$ in Equation~(\ref{eq:rhohat}) can be calculated analytically (see Appendix~\ref{sec:math}). Then, the {\it internal} MOND potential, $\Phi_{\rm in}$, which is the solution of the QUMOND equation (Equation~(\ref{eq:qumond})) {\it with the boundary condition $\mathbf{\nabla}\Phi_{\rm in}\rightarrow 0$ at infinity}, can be calculated through a direct three-dimensional integration as
\begin{equation}
    \Phi_{\rm in}(\mathbf{r})=-\int_0^\infty dr'\int_0^\pi d\theta' \int_0^{2\pi} d\phi' \frac{r'^2 \sin\theta' \hat{\rho}(r',\theta',\phi')}{|\mathbf{r}-\mathbf{r}'|},
 \label{eq:3Dint}
\end{equation}
for the density given by Equation~(\ref{eq:rhohat}). The EFE is the result of the fact that $\Phi_{\rm in}$ depends on $\mathbf{g}_{\rm N,ext}$, and, in particular, for any finite $\mathbf{g}_{\rm N,ext}$ it differs from the potential in the isolated case ($\mathbf{g}_{\rm N,ext}=0$).

\subsection{AQUAL algorithm}

To calculate the EFE in AQUAL, we need to solve the field equation (Equation~(\ref{eq:aqual})) for the given galaxy density distribution, with the boundary condition $-\mathbf\nabla\Phi\rightarrow \mathbf{g}_{\rm ext}$ at infinity, where $\mathbf{g}_{\rm ext}$ is the constant, external acceleration field in which the galaxy is falling. The internal dynamics is then governed by the acceleration field $-\mathbf{\nabla}\Phi-\mathbf{g}_{\rm ext}$. The fact that this internal field differs from the solution of the field equation with boundary condition $-\mathbf{\nabla}\Phi\rightarrow 0$ at infinity (the isolated case) is the expression of the EFE. (In Newtonian dynamics the two procedures give the same internal field.)

Except for highly symmetrical configurations, such as spherical ones, the nonlinear AQUAL field equation (Equation~(\ref{eq:aqual})) does not lend itself to an easy numerical solution such as Equation~(\ref{eq:3Dint}), especially so in the three-dimensional case required when the external field is not aligned with the disk axis. Here, we thus confine our AQUAL calculations to the axisymmetric configuration where the external field is parallel to the disk axis. We can then use the efficient and easy-to-implement two-dimensional algorithm proposed by \cite{Mil1986a}. 
In this algorithm, the modified Poisson Equation is recast as two coupled first-order differential equations for the vector field $\mathbf{U} \equiv \mu(|\mathbf{\nabla}\Phi|)\mathbf{\nabla}\Phi$, where $\Phi$ is the total MOND potential:
\begin{equation}
     \left\{ \begin{array}{rl}
           \mathbf{\nabla} \cdot \mathbf{U} & = 4\pi \rho , \\
           \mathbf{\nabla}\times [\nu(U)\mathbf{U}] & = 0 . \\
     \end{array} \right.
    \label{eq:aqualvec} 
\end{equation}
Here, $\nu(y)$ is related to $\mu(x)$ in Equation~(\ref{eq:aqual}) as in Equation~(\ref{eq:munu}), and the above choice we made in QUMOND of the `simple' IF $\nu(y)$ correspond to $\mu(x)=x/(1+x)$.

The reader is referred to \cite{Mil1986a} for the details of this algorithm. Here we only describe our numerical set-up. We consider a cylindrical space defined by $0 \le R \le R_{\rm max}$ and $-z_{\rm max} \le z \le z_{\rm max}$. We choose $R_{\rm max}=z_{\rm max}=500$ (in units of $r_{\rm M}$). We discretize the space as follows. For $R$, we let $R_{i+1}=R_{i}+\Delta R_{i+1}$ ($i=1,\cdots,N_R +1$) with $\Delta R_{i+1}=(1+\epsilon_R )\Delta R_i$ where $\epsilon_R = 0.05$ and $N_R=200$. In a similar way, we discretize $0\le z \le z_{\rm max}$ and $-z_{\rm max}\le z \le 0$ symmetrically with $\epsilon_z = 0.1$ and $N_z = 200$ (i.e.\ $N_z = 100$ on one side). Thus, we have $N_R \times N_z = 4\times 10^4$ cells for which components $U_R$ and $U_z$ are to be determined. (For a quicker and less heavy computation, one may choose $R_{\rm max}=z_{\rm max}=200$ and $N_R=N_z = 150$ without losing the numerical accuracy significantly.) Since we take for the MN scale lengths $b\le a$ and $a\sim r_{\rm M}$, our computational volume is much larger than the extent of the model galaxy.

The external field, of strength $e$, and directed in the $+z$ direction, strongly dominates at the chosen boundaries (the corresponding Newtonian field is $e_{\rm N}=e\mu(e)=e^2/(1+e)$). In this case, the analytic asymptotic expression derived in \cite{BM1984} is a very good approximation and can thus be used as the boundary condition. The detailed implementation is as described in \cite{Mil1986a}.

At the boundaries of $R$ we set
\begin{equation}
    U_R(R,\bar{z}_j)= \left. \frac{\partial \Phi_{\rm N}(R,\bar{z}_j)}{\partial R} \right|_{R=0,\hspace{0.1cm}R_{\rm max}},
    \label{eq:RBC}
\end{equation}
 where $\bar{z}_j\equiv (z_j + z_{j+1})/2$ is the $z$ coordinate at the midpoint of each cell at the boundaries. At the boundaries of $z$ we set 
 \begin{equation}
     \left. U_z(\bar{R}_i, z)=-e_{\rm N} + [1+\hat\mu(e)] \frac{\partial \Phi_{\rm N}(\bar{R}_i,z)}{\partial z} \right|_{z=\pm z_{\rm max}},
    \label{eq:zBC}
 \end{equation}
  where $\bar{R}_i\equiv (R_i + R_{i+1})/2$ is the $R$ coordinate at the midpoint of each cell at the boundaries. Note that, only $U_z$ (Equation~(\ref{eq:zBC})) has the important factor $1+\hat\mu(e)$ where $\hat\mu(e) \equiv d\ln\mu(x)/d\ln x|_{x=e} = 1/(1+e)$ (see section II(g) of \cite{Mil1986a}).
  
There are $N_{\rm dof}=2N_{\rm R}N_{\rm z}-N_{\rm R}-N_{\rm z}$ linear equations from Equation~(\ref{eq:aqualvec}) for the same number of unknowns, written as a vector $\mathbf{X}=\{U_{R}, U_z\}$, from the cells. When these equations are written in a matrix form ${\bf M}(\mathbf{X}) \mathbf{X}=\mathbf{S}$, where $\mathbf{S}$ is the source vector, most of the components of the $N_{\rm dof}\times N_{\rm dof}$ matrix ${\bf M}(\mathbf{X})$ are zero. We use the Python {\tt scipy.sparse.linalg} module to solve the matrix equation for $\mathbf{X}$. However, because the matrix is a function of $\mathbf{X}$, we solve the equation iteratively starting from an initial guess for $\mathbf{X}$ which is given by Equations~(\ref{eq:RBC}) and (\ref{eq:zBC}) at all (not just the boundaries) cells. The iteration is stopped when the maximum fractional change in the components of $\mathbf{X}$ is less than $10^{-6}$. 
Once the solution for $\mathbf{X}$ is found, we use the definition for $\mathbf{U}$ to solve for the components of the {\it internal} acceleration at the midpoint of each cell. Specifically, we solve numerically the coupled equations $U_R = -\mu(\sqrt{g_R^2+(g_z+e)^2})g_R$ and $U_z=-\mu(\sqrt{g_R^2+(g_z+e)^2})(g_z+e)$, where $g_R$ and $g_z$ are the $R$ and $z$ components of the internal acceleration. On the midplane of a disk, the magnitude of the radial acceleration is $g=-g_R$.

\section{Results} \label{sec:res}
The EFE modifies the internal field of the galaxy everywhere. But in this paper we are concerned with the signature of the EFE in the rotation curves. All our results will thus refer to the modifications of the internal accelerations in the plane of the galaxy. Instead of showing the affected (radial) acceleration as a function of galactic radius for each model, we shall present our results as the modified ``mass-discrepancy-acceleration relation'' aka the ``radial acceleration relation (RAR)''. For the models in which the external field is inclined to the disk axis -- where the radial acceleration is a function of both radius and azimuthal angle -- we show the azimuth-averaged acceleration at a given radius, as a function of the (azimuth-independent) Newtonian acceleration at this radius.

\subsection{External field parallel to the rotation axis} \label{sec:paral}

\begin{figure*}
  \centering
  \includegraphics[width=1.\linewidth]{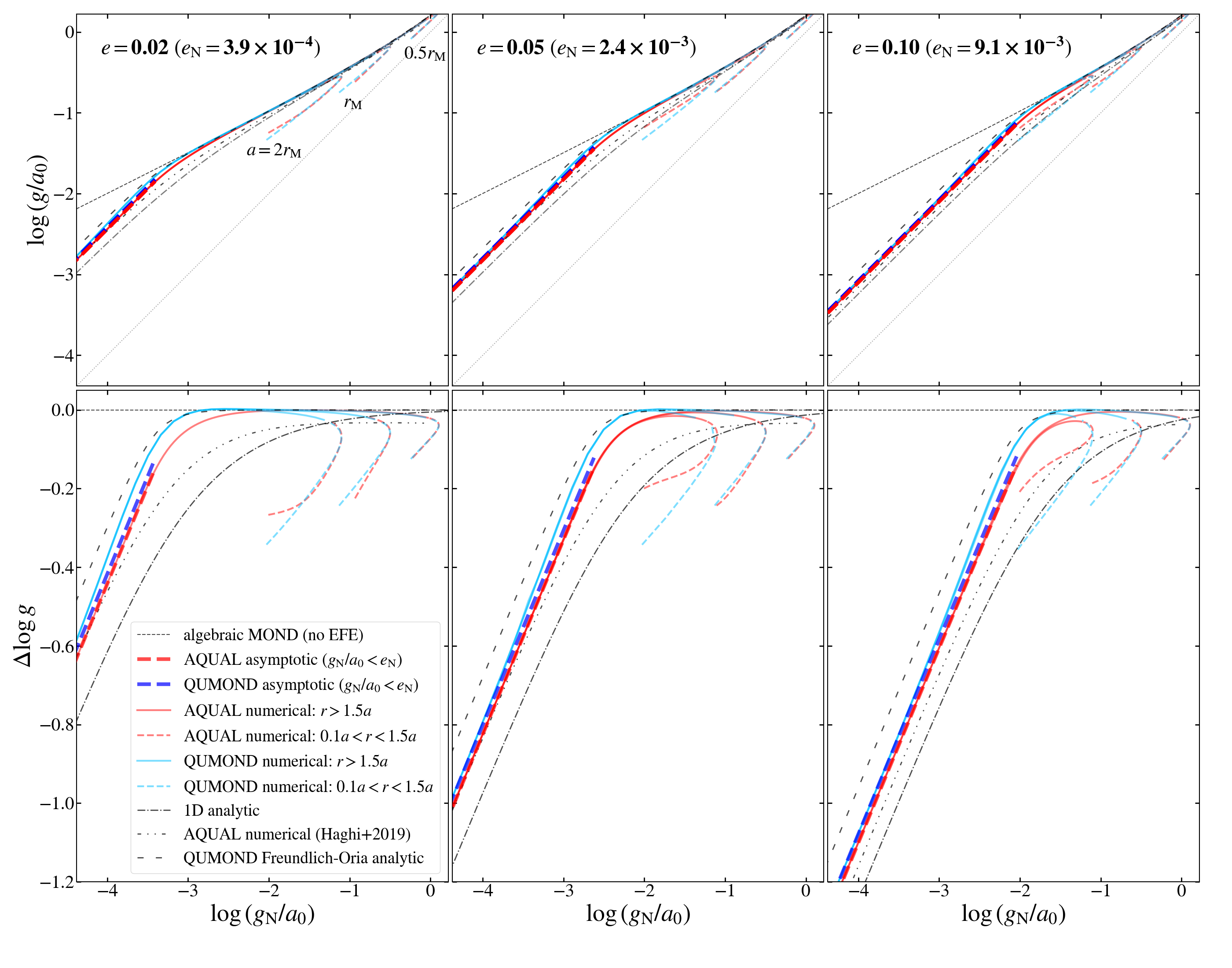}
    \vspace{-0.5truecm}  
    \caption{\small 
    \textbf{External field dependent radial acceleration relation.}
    Top panels show Numerical results for the radial acceleration in the midplane of a disk under a constant \emph{external field parallel to the rotation axis}. The considered external field strengths in units of $a_0$ are $e=0.02$, $0.05$ and $0.1$ (the Newtonian external field is $e_{\rm N}=e^2/(1+e)$). The black dashed curve represents the algebraic MOND relation without EFE while the gray dotted line is the Newtonian relation $g=g_N$. The bottom panels show the relative radial acceleration with respect to the algebraic MOND relation without EFE. Three cases of disk scale length in the MN model are considered: $a=2r_{\rm M}$, $r_{\rm M}$, and $0.5 r_{\rm M}$ where $r_{\rm M}=\sqrt{GM/a_0}$. The model with a higher $a$ corresponds to galaxies of lower surface density such as dwarf galaxies and LSBs. QUMOND and AQUAL results are respectively denoted by light blue and red colored curves: the outer ($r>1.5a$) and the inner ($r<1.5a$) parts are distinguished by solid and dashed curves. The outer parts agree well with the corresponding analytic asymptotic limits from Equations~(\ref{eq:pointa}) and (\ref{eq:pointq}), up to $g_{\rm N}/a_0=e_{\rm N}$ in the case of AQUAL.  Our results are also compared with literature numerical or approximate results. The QUMOND results match recent simulation or analytic results \citep{Zon2021,Freund2021,Oria2021}. However, the AQUAL results show differences from previously published simulation or analytic results \citep{FM2012,Haghi2019}.} 
   \label{efe_paral}
\end{figure*} 

We first consider disk systems for the case that the external field is parallel to the rotation axis as this axisymmetric case allows numerical solutions in both AQUAL and QUMOND based on our algorithms. For the disk geometry, we take $b/a=0.1$ (effects of thicker disks are considered later) and consider a likely range of $a$ (in units of $r_{\rm M}$) motivated by observed disk galaxies. Massive and high surface density galaxies such as the Milky Way have $a\la 0.5$ while low surface density galaxies such as LSBs and dwarf galaxies have $a>1$ \citep{Lel2016}. We also consider values of $e$ motivated by the observed environments of SPARC galaxies \citep{Chae2020,Chae2021}. They report that external fields typically are $0.02\la e \la 0.1$. 

Figure~\ref{efe_paral} shows the radial acceleration as a function of the Newtonian radial acceleration in the midplane of the disk. We exhibit three cases of $a=0.5$, 1, and 2 for three cases of $e=0.02$, $0.05$, and $0.1$. As expected, the EFE starts to decrease the radial acceleration far away from the galactic center ($r\gg a$) where Newtonian accelerations are very low $g_{\rm N} \la 10^{-11}$~m~s$^{-2}$. Our numerical results match accurately the corresponding analytic asymptotic limits of AQUAL and QUMOND (Equations~(\ref{eq:pointa}) and (\ref{eq:pointq})). Note that published N-body based numerical simulation results in AQUAL \citep{Haghi2019} or QUMOND \citep{Oria2021} do not match the asymptotic limits precisely probably due to the limits of numerical precision.

From Figure~\ref{efe_paral} we find the following aspects of the EFE in disk galaxies.
\begin{enumerate}
    \item The EFE in both theories is, as expected from the one-dimensional approximation, weaker than the approximate one-dimensional model by \cite{FM2012}. It is also weaker than the simulation results for velocity dispersions in spherical systems by \cite{Haghi2019}. Consequently, for the currently observed outskirts of disk galaxies with $g_{\rm N}/a_0\la 0.1$ the EFE can start to be noticeable near $e_{\rm N}\sim 0.01$ (i.e. $e\sim 0.1$). 
    \item The EFE is stronger in AQUAL than in QUMOND. In principle, RCs of galaxies under a relatively strong external field can distinguish two theories at $g_{\rm N}/a_0 \la 0.1$. 
    \item The inner region $r\la 1.5 a$, where the RC is rising with radius, the computed accelerations deviate from both the radial acceleration without EFE as calculated from the algebraic formulation (black dashed curve) and the Newtonian radial acceleration. Most of the former departure is not due to an EFE but to the difference in formulations, as it is present even when external field is weak and is more prominent when $a/r_{\rm M}$ is large. This difference in formulations can matter most for the RCs of dwarf galaxies and LSBs with $a > r_{\rm M}$. For high-density galaxies with $a\ll r_{\rm M}$, this deviation is expected to be very weak and not observable because in the Newtonian limit all formulations tend to the Newtonian behavior.
\end{enumerate}

\begin{figure}
  \centering
  \includegraphics[width=1.0\linewidth]{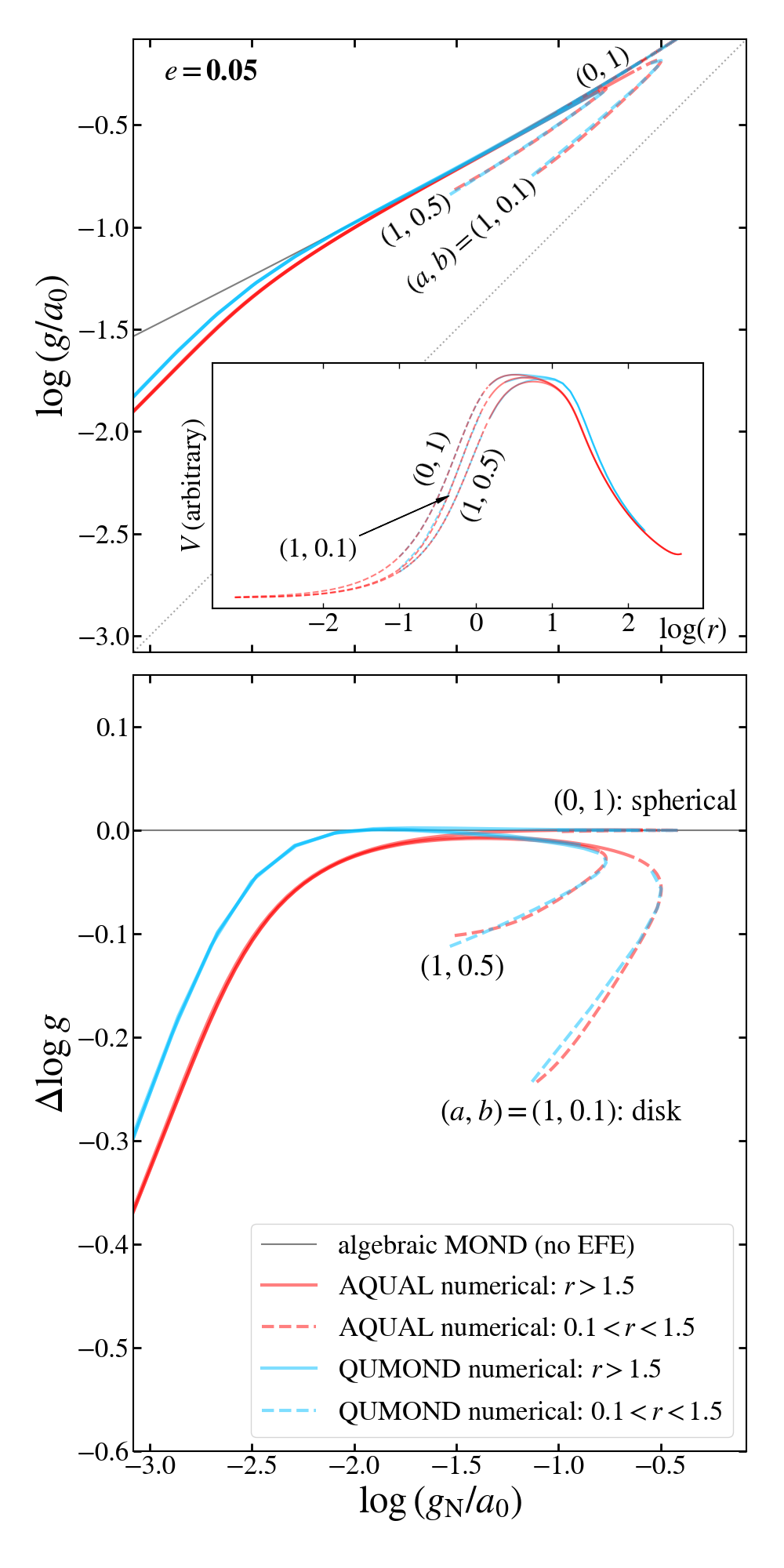}
    \vspace{-0.5truecm}  
    \caption{\small 
    \textbf{Effect of disk thickness on the RAR under an external field.}
  In the inner part ($r<1.5a$) where the rotation curve is rising with radius, the RAR depends critically on the thickness of the disk. The inset of the upper panel shows the rotation curve in a logarithmic scale. The RAR in the inner part deviates more from that of the algebraic MOND without EFE for a thinner disk. For a spherical system, the inner and outer parts fall on the line of the algebraic MOND without EFE in the high acceleration limit. The outer part is little affected by the disk thickness.
  }
   \label{efe_thick}
\end{figure}

The last kind of departure above has not been observed in spherical systems. This suggests that the behavior for $r < 1.5a$ will depend on the thickness of the disk because the more spherical the system is the more similar the different formulations become. Indeed, Figure~\ref{efe_thick} shows that there is no deviation in the inner region for the spherical system while the deviation is weaker for a moderately flattened system. Thus, the inner region $r < 1.5a$ depends critically on the intrinsic shape of the system (see also \cite{angus12} who describe such effects of the disk thickness). However, note also that the outer part $r > 1.5a$ does not depend on the intrinsic shape but follows a common EFE-affected RAR.

The departure of the inner part from the algebraic MOND relation, Equation~(\ref{eq:algeb}), without EFE (which is exactly valid for the spherical system) is a direct consequence of the disk geometry. Then, how does it depend on the external field? Comparison of three columns of Figure~\ref{efe_paral} indicates that external field has some effects on its shape in the radial acceleration space.

\begin{figure}
  \centering
  \includegraphics[width=1.0\linewidth]{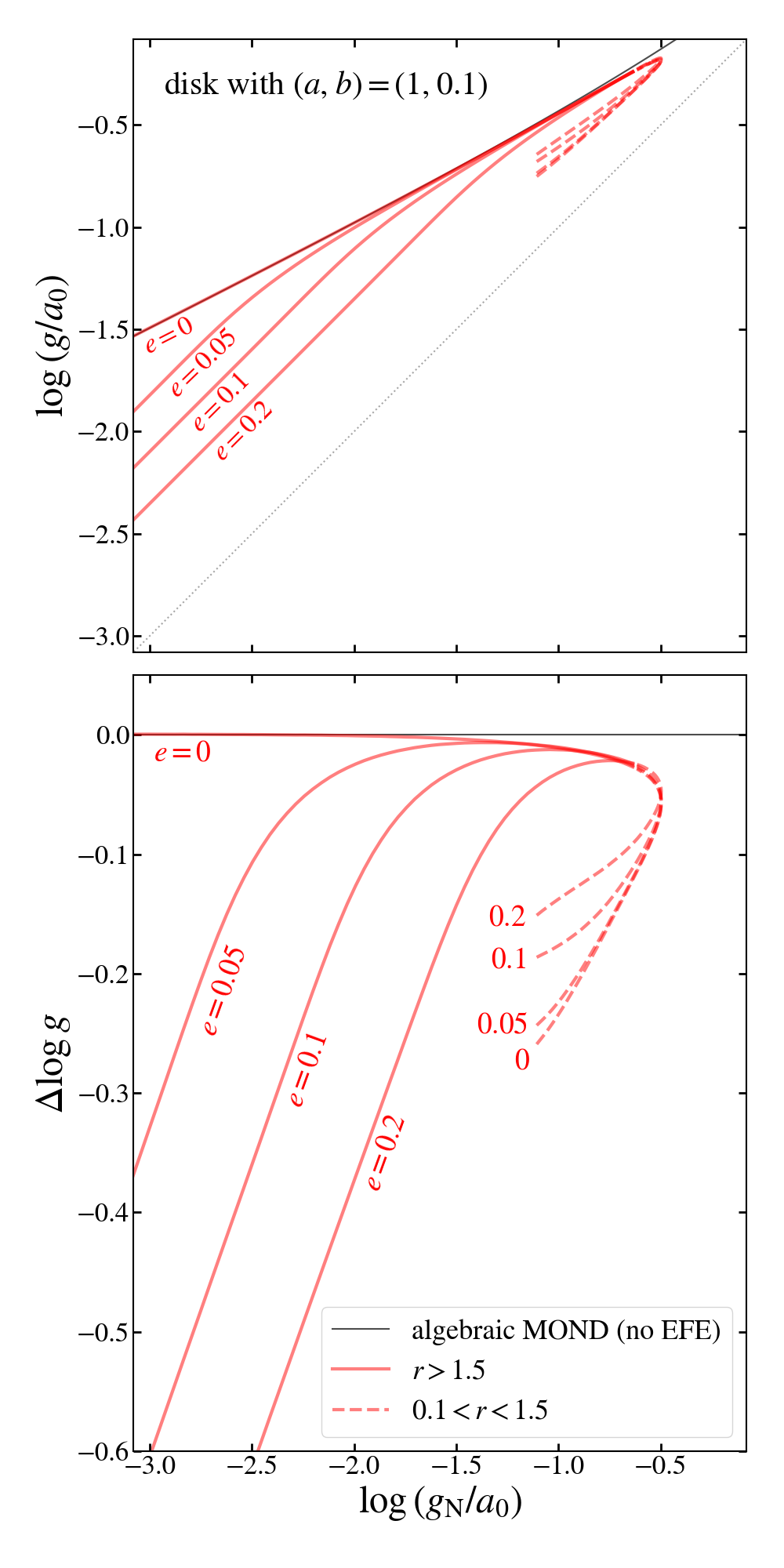}
    \vspace{-0.5truecm}  
    \caption{\small 
    \textbf{Variation of a disk RAR across a broad range of external field.}
  Numerical results for the AQUAL model are shown for a broad range of $0\le e \le 0.2$. As the external field is varied, the inner part ($r<1.5a$) has a rather mild dependence on $e$ while the outer part is dramatically varied. Note that even when $e=0$ (no external field), the inner part deviates from the algebraic relation, Equation~(\ref{eq:algeb}).
  }
   \label{efe_eval}
\end{figure}

To see the EFE on the inner part, we consider a broader range of $e$ from $0$--$0.2$. Figure~\ref{efe_eval} shows the results for the AQUAL model. (We do not consider the QUMOND case because the numerical integration of Equation~(\ref{eq:3Dint}) becomes extremely slow as $e\rightarrow 0$.) The inner part deviates from the algebraic MOND always even when $e=0$ while the outer part of the RAR deviates only when $e\ne 0$. The inner part has a rather mild dependence on $e$ compared with the departure of the disk when $e=0$ from the algebraic MOND RAR. Perhaps, counter-intuitively the inner part gets closer to the algebraic MOND RAR when $e$ gets larger. 

\subsection{External field tilted from the rotation axis} \label{sec:tilt}

We now consider the general case that the external field is tilted from the rotation axis. In this case the geometry is no longer axisymmetric and even a warping of the disk is expected \citep{BM2000}. For this three-dimensional geometry we have a numerical solution for the radial acceleration as a function of azimuthal angle only for the QUMOND algorithm. For the AQUAL case we rather consider the spherical geometry and analyze circular orbits whose axes are tilted from the $z$-axis. This approximation should suffice for the outer part $r > a$ as it is not sensitive to the intrinsic shape as shown in Figure~\ref{efe_thick}.

\begin{figure}
  \centering
  \includegraphics[width=1.0\linewidth]{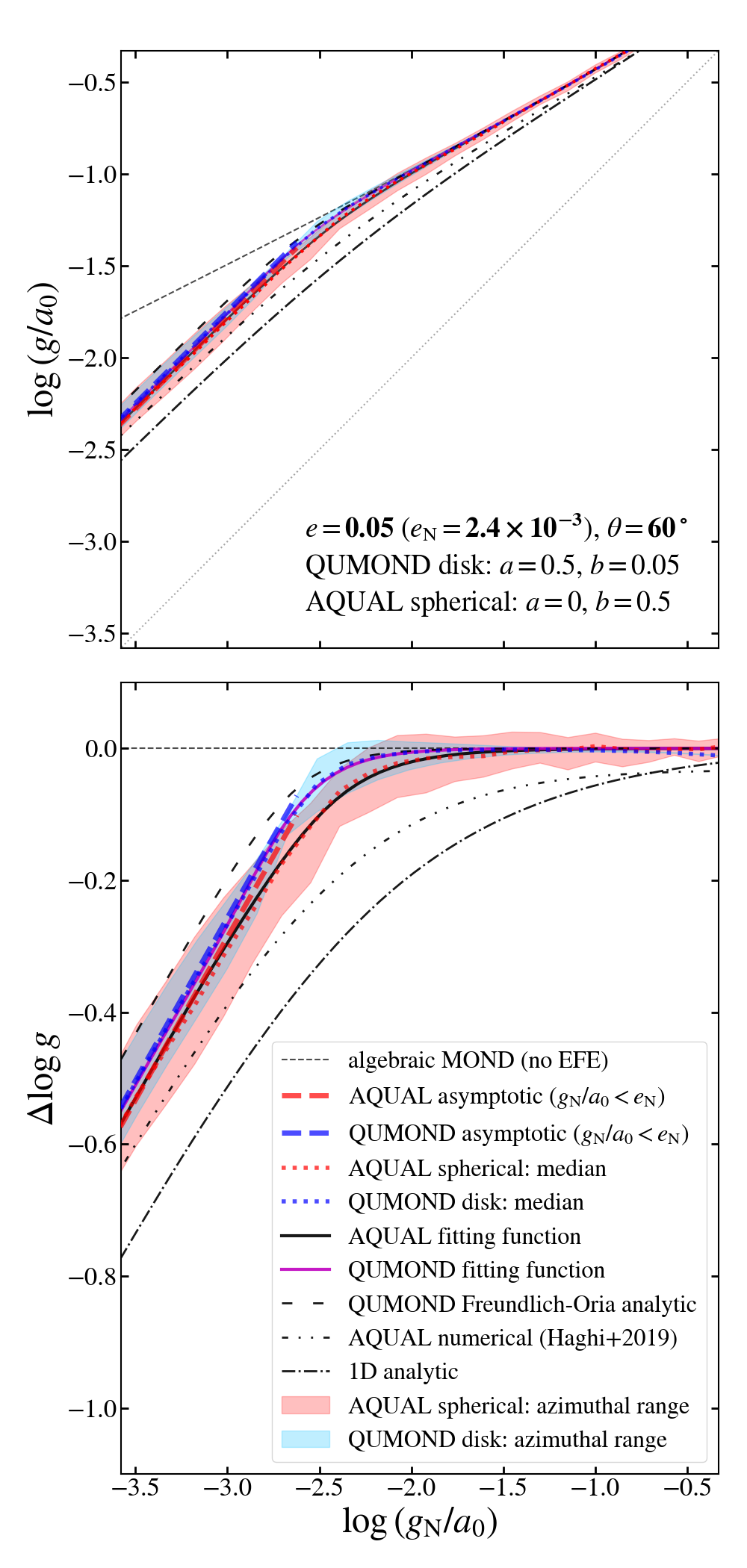}
    \vspace{-0.8truecm}  
    \caption{\small 
    \textbf{Effect of the tilt of the external field for the outer part.}
  There is a scatter of the radial acceleration along the azimuthal angle when the external field is tilted from the rotation axis. The tilt angle of $\theta=60^\circ$ corresponds to the average of random orientations in a three-dimensional space. For the QUMOND case, a disk model shown in Figure~\ref{efe_paral} is considered. For the AQUAL case, a tilted circular orbit in a spherical model is considered instead because the AQUAL code does not allow a three-dimensional calculation. The QUMOND median result is well described by Equation~(\ref{eq:qumondfit}) (magenta curve). The AQUAL median result is well described by Equation~(\ref{eq:aqualfit}) (black curve). The median results agree well with the asymptotic limits for $g_{\rm N}\la 0.5 e_{\rm N}a_0$. The \cite{Haghi2019} numerical fitting function is adjusted here to the simple IF from the standard IF. }
   \label{efe_tilt}
\end{figure}

Figure~\ref{efe_tilt} shows an azimuthal scatter of the radial acceleration for the case that the tilt angle is $\theta=60^\circ$ (the average of random orientations). The scatter is larger in the AQUAL solution than the QUMOND, but this comparison is not fully valid because a spherical geometry is used in the AQUAL case. The azimuthal scatter is anyhow larger than the difference between the median relations for the two theories (blue and red dashed curves in Figure~\ref{efe_tilt}). 

The QUMOND numerical median relation for the outer part agrees well with the approximate analytic expression for the RAR proposed by \cite{Zon2021} based on \cite{BZ2018}:
\begin{equation}
    \left. g\right|_\text{Q} = g_{\rm N} \nu(y_1) \left[1+\tanh\left(\frac{0.825e_{\rm N}}{y}\right)^{3.7} \frac{\hat{\nu}(y_1)}{3} \right],
    \label{eq:qumondfit}
\end{equation}
where $y=g_{\rm N}/a_0$ and $y_1 \equiv \sqrt{y^2+e_{\rm N}^2}$. Equation~(\ref{eq:qumondfit}) was actually obtained for an external field parallel to the rotation axis but still matches reasonably well our median relation when the disk is tilted.  Note that the analytic formula obtained by \cite{Freund2021} from averaging the QUMOND algebraic relation over a sphere is similar to this function but deviates in the asymptotic limit, as is also seen in \cite{Oria2021}. 

\begin{figure}
  \centering
  \includegraphics[width=1.0\linewidth]{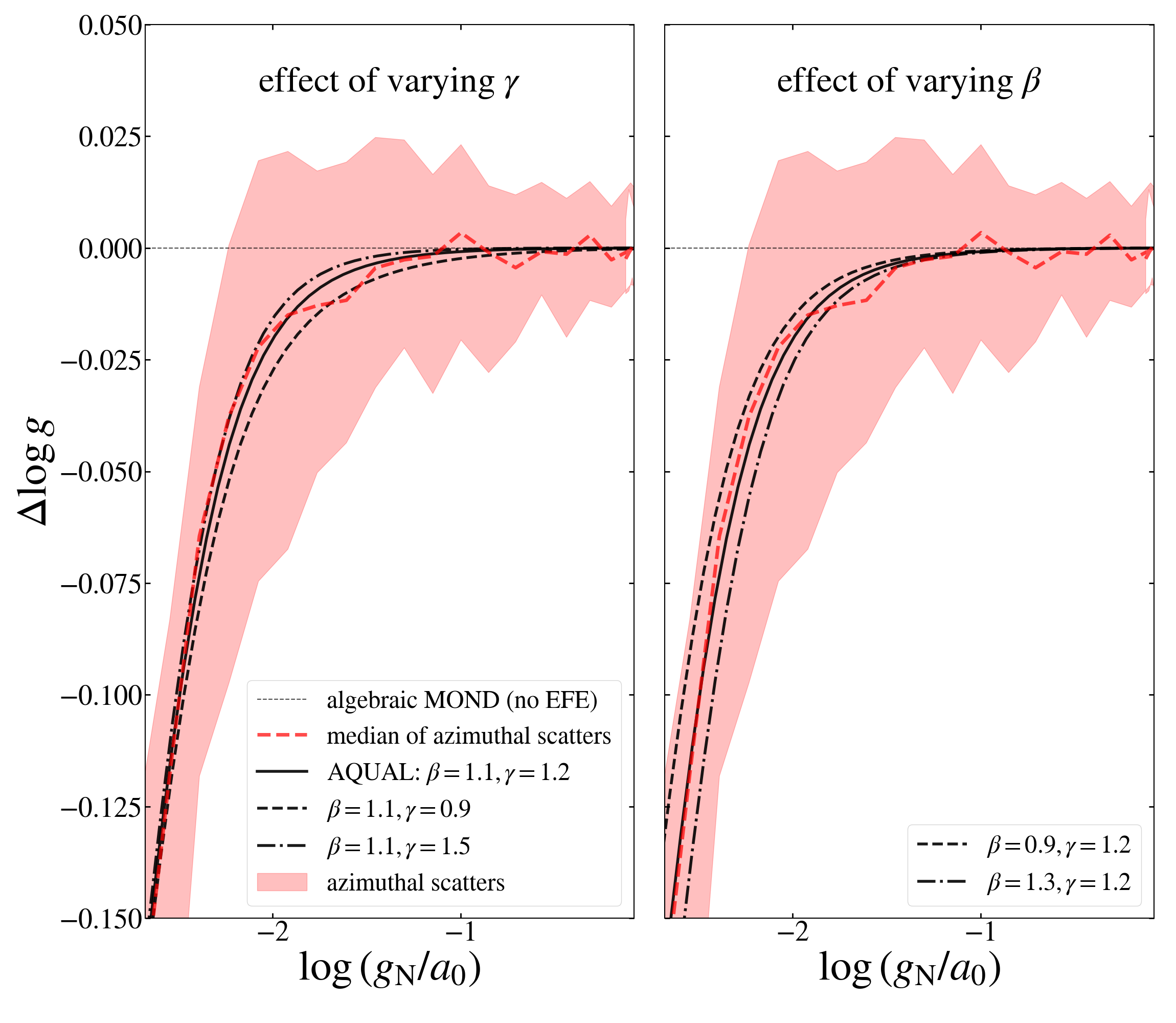}
    \vspace{-0.4truecm}  
    \caption{\small 
    \textbf{Effects of varying parameters of the AQUAL fitting function.}
  Parameters $\beta$ and $\gamma$ of the AQUAL fitting function (Equation~(\ref{eq:aqualfit})) are varied from the curve shown in Figure~\ref{efe_tilt}. }
   \label{AQUALvaried}
\end{figure}

We find that the AQUAL numerical results for the outer part can be described to a good approximation by the following functional form:
\begin{equation}
     \left. g\right|_\text{A} = g_{\rm N} \nu(y_\beta) \left[1+\tanh\left(\frac{\beta e_{\rm N}}{y}\right)^{\gamma} \frac{\hat{\nu}(y_\beta)}{3} \right],
    \label{eq:aqualfit}
\end{equation}
where $y_\beta \equiv \sqrt{y^2+(\beta e_{\rm N})^2}$. The AQUAL median relation is well described by $\beta=1.1$ and $\gamma=1.2$ as shown in Figure~\ref{efe_tilt}. This parameter choice is designed to describe best the acceleration regime of the outer parts, but not below $\sim 10^{-13}$~m~s$^{-2}$. This choice also approximates well numerical results for the disk under an external field parallel to the rotation axis. For lower accelerations, $\la 10^{-13}$~m~s$^{-2}$, or for numerical results based on IFs different from the `simple' one, values of $\beta$ and/or $\gamma$ may be adjusted up to $\pm 0.2$. Effects of varying these parameters are exhibited in Figure~\ref{AQUALvaried}. The above fitting functions and other functions/formulas from the literature are summarized in Table~\ref{tab:formulae}.

 \begin{table*}
\caption{Summary of Fitting Functions of the EFE-dependent RAR for the Outer Part of Rotation Curves}\label{tab:formulae}
\begin{center}
  \begin{tabular}{llll}
  \hline
 case & theory  &  $\frac{g_{\rm MOND}}{g_{\rm N}}(y,e)$ for $y\equiv \frac{g_{\rm N}}{a_0}$ and $e\equiv \frac{g_{\rm ext}}{a_0}$ (or $e_{\rm N}\equiv \frac{g_{\rm N,ext}}{a_0}$) & references \\
 \hline
 (0) algebraic MOND (no EFE)  &  generic & $\nu(y)$ ($e=0$) &  M83  \\
 (1) asymptotic limit & AQUAL &  $\frac{1}{\mu(e)} \left( 1+\hat\mu(e) - \hat\mu(e) \frac{\sin^2\theta}{2} \right)^{-1/2} $     &    M10, C21   \\
 (2) asymptotic limit & QUMOND &    $\nu(e_{\rm N}) \left(1+\frac{\hat{\nu}(e_{\rm N})}{2} - \hat{\nu}(e_{\rm N}) \frac{\sin^2\theta}{4}\right) $  &   M10, C21      \\
 (3) 1D analytic & generic  & $ \frac{1}{2}+\sqrt{\left(\frac{1}{2}-e\frac{A_e}{y}\right)^2+\frac{B_e}{y}}-e\frac{A_e }{y} $  & FM12, C21  \\
 (4) Freundlich-Oria analytic  & QUMOND  &   $\nu\left({\rm min}\left[y+\frac{e_{\rm N}^2}{3y},e_{\rm N}+\frac{y^2}{3e_{\rm N}}\right]\right)$ &  F21, O21  \\
 (5) QUMOND numerical  &  QUMOND  &  $\nu(y_1) \left[1+\tanh\left(\frac{0.825e_{\rm N}}{y}\right)^{3.7} \frac{\hat{\nu}(y_1)}{3} \right]$ &  Z21 \\
 (6) AQUAL numerical  & AQUAL & $\nu(y_\beta) \left[1+\tanh\left(\frac{\beta e_{\rm N}}{y}\right)^{\gamma} \frac{\hat{\nu}(y_\beta)}{3} \right]$   &  this work  \\
 (7) AQUAL numerical  & AQUAL   & $\frac{2\sqrt{2}}{3\sqrt{y}}[1+0.56\exp(3.02\log_{10}(y))]^{0.368} 10^{2F_e}$ &  H19 \\
 \hline
  \end{tabular}
 
  Note. (0) This is the MOND IF itself: e.g., $\nu(y)=1/2+\sqrt{1/4+1/y}$ (the `simple' IF),  $\nu(y)=\sqrt{1/2+\sqrt{1/4+1/y^2}}$ (the `standard' IF). (1) $\mu(x)$ is the inverse form of $\nu(y)$: e.g., $\mu(x)=x/(1+x)$ (simple), $\mu(x)=x/\sqrt{1+x^2}$ (standard). $\hat{\mu}(x)\equiv d\ln\mu(x)/d\ln x$. $\theta$ is the angle between the $z$-axis and the external field $\mathbf{g}_{\rm ext}$. (2) $\hat{\nu}(y)\equiv d\ln \nu(y)/d\ln y = -\hat{\mu}(x)/(1+\hat{\mu}(x))$ where $x=\nu(y)y$. $\theta$ is the angle between the $z$-axis and the Newtonian external field $\mathbf{g}_{\rm N,ext}$. (3) This relation is for the simple IF with $A_e \equiv (1+e/2)/(1+e)$ and $B_e\equiv 1+e$. (4) min($a$, $b$) refers to the minimum of $a$ and $b$. (5) \& (6) These numerical results are for the simple IF with $y_\beta \equiv \sqrt{y^2+(\beta e_{\rm N})^2}$. The choice $\beta=1.1$ \& $\gamma=1.2$ works best in the acceleration regime stronger than $\sim 10^{-13}$~m~s$^{-2}$. (7) This numerical result is for velocity dispersions in a spherical system assuming the standard IF. The following definition is used: $F_e\equiv -(A'_e/4)[\ln(\exp(-\log_{10}(y)/A'_e)+B'_e)+C'_e]$ where $A'_e\equiv 5.3/[10.56+(\log_{10}(e)+2)^{3.22}]$, $B'_e\equiv 10^{-[1.65\log_{10}(e)+0.0065]}$, $C'_e\equiv 3.788 \log_{10}(e)+0.006$.
        
  References. M83 - \cite{Mil1983}, M10 - \cite{Mil2010}, C21 - \cite{Chae2021}, FM12 - \cite{FM2012}, F21 - \cite{Freund2021}, O21 - \cite{Oria2021},
  Z21 - \cite{Zon2021}, H19 - \cite{Haghi2019}
 \newline 
\end{center}
\end{table*}

For a larger tilt angle than $\theta=60^\circ$, the azimuthal scatter becomes larger, but the median is little changed. Thus, our QUMOND results confirm the QUMOND prediction \citep{Zon2021,Oria2021} that EFE is very weak in the outer quasi-flat part of the RC. However, while AQUAL predicts stronger EFE than QUMOND, our AQUAL results predict weaker EFE than the \cite{Haghi2019} results in the outer part. 

\subsection{Examples: HSB and LSB Galaxies under external fields}  \label{sec:examples}

For the purpose of illustration (rather than precise modeling) of the prediction of our numerical results on the RCs of real galaxies, we consider simple mass models of two prototype galaxies under relatively strong external fields: the high surface brightness (HSB) galaxy NGC~5055 and the low surface brightness (LSB) galaxy F568-V1 from the SPARC database \citep{Lel2016}.

\begin{figure}
  \centering
  \includegraphics[width=1.0\linewidth]{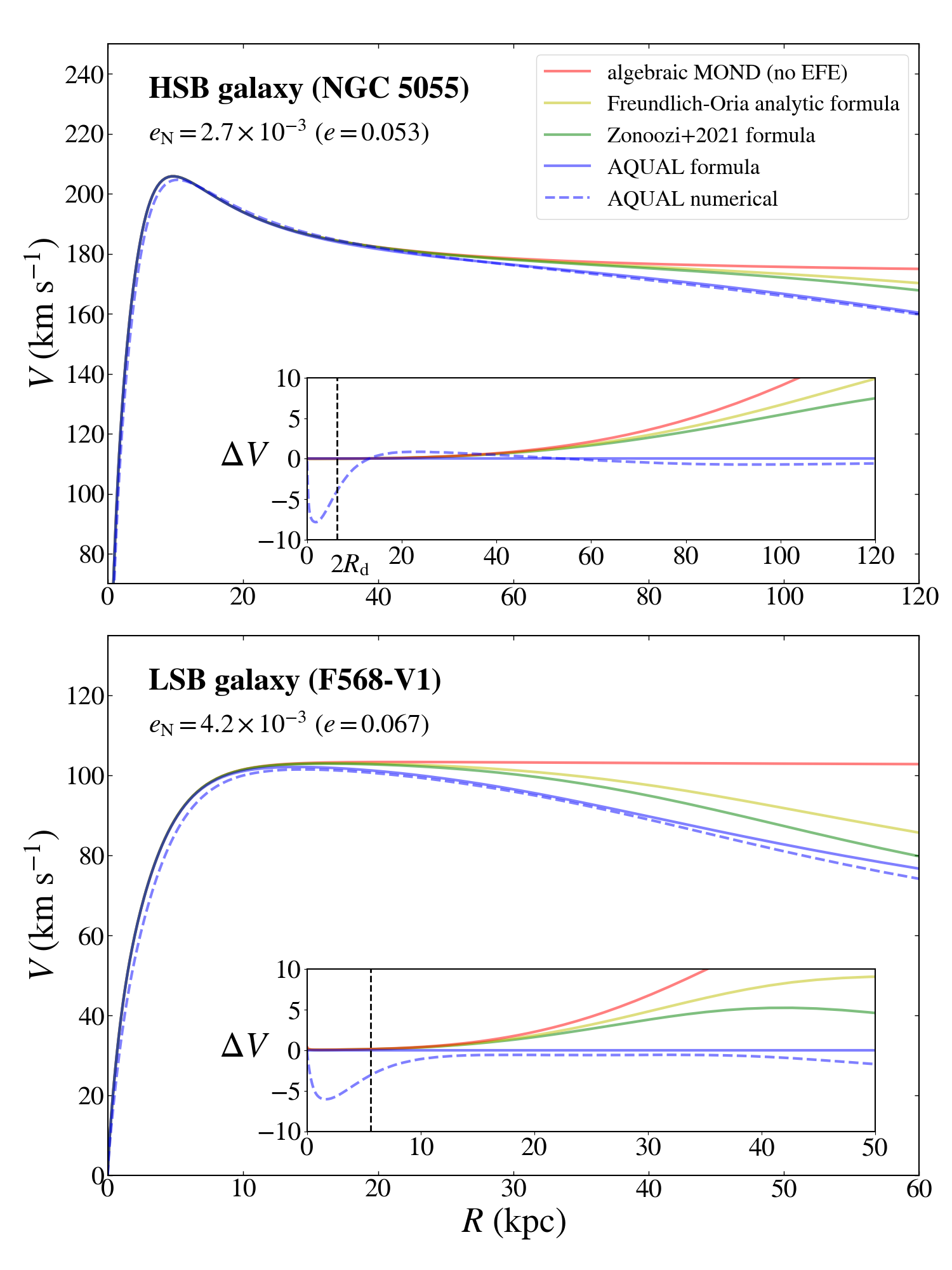}
    \vspace{-0.4truecm}  
    \caption{\small 
    \textbf{Simple models of two galaxies.}
  For an HSB galaxy NGC~5055 and an LSB galaxy F568-V1, both of which are under relatively strong external fields, simple mass models as described in Section~\ref{sec:examples} are considered. AQUAL numerical predictions (blue dashed curves) are compared with some of the fitting functions listed in Table~\ref{tab:formulae}. For the same mass model and the same external field, the AQUAL theory (blue curve) predicts a stronger deviation from the algebraic MOND flat RC (red line) than the QUMOND theory (yellow, green curves). }
   \label{examples}
\end{figure}

NGC~5055 (the ``sunflower'' galaxy) is one of the best examples of giant spiral galaxies with relatively well-measured extended RCs. This galaxy was also modeled by \cite{Oria2021} using their QUMOND simulation results. The baryonic mass distribution of this galaxy is dominated by the stellar disk. Here we model this galaxy using an approximate exponential disk composed of three MN models (`MN3' model in Figure~\ref{MN_expdisk}) ignoring the subdominant gas disk. As in \cite{Oria2021}, we assume a total baryonic mass of $M_{\rm bar}=5.48\times 10^{10}{\rm M}_\odot$ for which the MOND radius is $r_{\rm M}=\sqrt{GM_{\rm bar}/a_0}=7.98$~kpc for $a_0=1.20\times 10^{-10}$~m~s$^{-2}$. For the exponential disk scale length we take $R_{\rm d}=3.20~{\rm kpc}\approx 0.40r_{\rm M}$ and a common MN thickness of $0.040 r_{\rm M}$. For the Newtonian circular speed $V_{\rm N}(R)$ at a cylindrical radius $R$ on the midplane of the disk, the MOND predicts a circular speed given by
\begin{equation}
  V_{\rm MOND}(R) = \sqrt{\frac{g_{\rm MOND}[g_{\rm N}(R)/a_0,e_{\rm N}]}{g_{\rm N}(R)}} V_{\rm N}(R),
  \end{equation}
where $g_{\rm MOND}(y,e_{\rm N})/g_{\rm N}$ is gotten from the AQUAL numerical calculation for an assumed external field of $e_{\rm N}=2.7\times 10^{-3}$ (i.e.\ $e=0.053$). The numerical prediction is then compared with the predictions of some models in Table~\ref{tab:formulae}.

The outer part of the predicted RC of NGC~5055 agrees well with the AQUAL fitting function and starts to deviate from the algebraic MOND (red) flat line near $R\approx 40$~kpc somewhat earlier than the QUMOND deviation occurring near $R\approx 65$~kpc. This is an intriguing difference that can have important consequences in actual precise modeling of the galaxy to fit the observed RC. Because the QUMOND prediction can be made to match roughly the AQUAL prediction by invoking a higher external field, QUMOND is generically expected to require a higher external field than AQUAL to fit the same \emph{declining} RC. In passing, we note that the AQUAL prediction based on this simple model appears to match the observed RC (see Figure~2 of \citealt{Chae2020}) somewhat better than the QUMOND prediction.

The inner part ($\la 2R_{\rm d}$) of the predicted RC of NGC~5055 shows a deviation from the algebraic MOND consistent with the numerical results shown in Figures~\ref{efe_paral},~\ref{efe_thick}, and~\ref{efe_eval}. Note also that all fitting functions applicable only to the outer part converge to the algebraic MOND by design. Thus, if any of the fitting functions listed in Table~\ref{tab:formulae} is used to fit an observed RC, the inner rising part should not be used.

F568-V1 is an LSB galaxy whose baryonic mass is dominated by an extended gas disk rather than the smaller stellar disk. We assume a total baryonic mass of $M_{\rm bar}=6.4\times 10^{9}{\rm M}_\odot$ with the corresponding MOND radius of $r_{\rm M}=2.8$~kpc. The gas disk is assumed to occupy 52\% of the baryonic mass. For the stellar disk, we use the MN3 exponential disk model with $R_{\rm d}=2.84~{\rm kpc}\approx r_{\rm M}$ and $b=0.1r_{\rm M}$ while for the gas disk, we use an MN model with $a=4r_{\rm M}$ and $b=0.4r_{\rm M}$. For the external field we take $e_{\rm N}=4.2\times 10^{-3}$ (i.e.\ $e=0.067$) from \cite{Chae2021}.

As expected from Figure~\ref{efe_paral}, the predicted inner RC for F568-V1, which has a larger $R_{\rm d}/r_{\rm M}$ ratio than NGC~5055, deviates more clearly from the algebraic MOND curve. The maximum deviation in the inner part amounts to $5.9$\% of the maximum circular speed while it is $3.8$\% in NGC~5055. As in NGC~5055, the outer part is adequately described by the AQUAL fitting function and shows a difference from the QUMOND fitting functions. However, in a very deep MOND regime the AQUAL fitting function starts to deviate from the numerical result because the parameter choice ($\beta=1.1$ \& $\gamma=1.2$) was optimized for an acceleration regime $\ga 10^{-13}$~m~s$^{-2}$.

\section{Discussion} \label{sec:disc}

We have obtained numerical results for the EFE on the RCs of disk galaxies using the QUMOND and AQUAL theories. We have investigated both the quasi-flat outer part ($r>1.5a$) and the rising inner part ($r<1.5a$) of the RCs. For the outer part we find that our QUMOND results confirm recently published numerical results \citep{Zon2021,Oria2021}. Our results are particularly well described by the fitting function proposed by \cite{Zon2021} (Equation~(\ref{eq:qumondfit})). This function also matches accurately the QUMOND analytic asymptotic limit. AQUAL results imply stronger EFE than QUMOND. We propose a new fitting function (Equation~(\ref{eq:aqualfit})) which is more accurate than published approximate AQUAL results. Our AQUAL results represent the first numerical results ever reported that match accurately the AQUAL analytic asymptotic limit. 

Equations~(\ref{eq:qumondfit}) and (\ref{eq:aqualfit}) summarized here can be used to study the EFE in the outer quasi-flat parts of observed RCs. Also, because there is some difference between the two functions at low accelerations $g_{\rm N}/a_0 \la 0.1$ when the external field is relatively strong $e \ga 0.1$, a statistical analysis of good-quality RCs may discriminate two theories eventually.

Our results (Figure~\ref{efe_paral} and \ref{efe_eval}) show that the RAR of the inner rising part of the RCs for flattened systems clearly deviates from that of the algebraic RAR without EFE, in agreement with the previous results mentioned above. The downward deviation of the inner part is driven by the disk geometry, i.e.\ the flattening. In other words, there is no deviation in the inner part when there is no flattening. External fields can further modify the inner part RAR. However, for typical external fields ($e \la 0.2$) the effect is rather minor compared with that of flattening (Figure~\ref{efe_eval}).

This rather peculiar behavior of the inner rising part of a disk is another prediction of MOND (AQUAL and QUMOND alike: see Figure~\ref{efe_paral}). This is a consequence of the non-linear nature of MOND field equations. The behavior of the inner part in the RAR space (Figure~\ref{efe_paral}) is more important for galaxies with $a>r_{\rm M}$. These are galaxies with low surface mass densities and thus smaller $r_{\rm M}$ compared with galaxy sizes. For galaxies with a small $a$ ($\ll r_{\rm M}$), i.e.\ galaxies with high surface densities, the inner part  ends at a high acceleration $g_{\rm N}\ga a_0$ so that there is little room for an observable deviation. Those galaxies often have bulges which increase the inner gravity. 

Dwarf galaxies and LSBs typically have $a>r_{\rm M}$ and the observed RCs significantly contain (or are dominated by) rising RCs under weak radial acceleration ($\la 10^{-10.5}$~m~s$^{-2}$) for $r<1.5a$. When RCs are statistically analyzed to study the EFE, it will be important not to mix these inner RCs with outer RCs. As Figure~\ref{efe_paral} shows, the inner RCs can mimic outer RCs with high values of $e$.  {(In this context, \citealt{pl20} have made an attempt to distinguish between the algebraic MOND relation and modified gravity theories, from the RCs of dwarf galaxies, based on the criterion suggested in \citealt{Mil2012b}.)}

Our results can be used to refine the EFE analyses of SPARC galaxies \citep{Chae2020,Chae2021}. The new fitting function (Equation~(\ref{eq:aqualfit})) can be used to revise $e$ values by fitting the outer quasi-flat RCs. Our results on the inner RCs can be used to reinterpret the RCs of dwarf galaxies and LSBs.

The difference between QUMOND and AQUAL results and the fact that the `ultimate' MOND theory will probably differ from both, imply that there are some theoretical uncertainties in the quantitative prediction of EFE. Moreover, relativistic MOND theories are still under development (see, e.g., \citealt{SZ2021}). Thus, comparison of results such as  Equation~(\ref{eq:aqualfit}) with observations may help discriminate between MOND theories.

\vspace{0.3in}

\acknowledgements

 K.-H.C. thanks I. Banik for discussions. We thank the anonymous reviewer for useful suggestions to improve the presentation. This research was supported by the National Research Foundation of Korea (NRF) grant funded by the Korea government(MSIT) (No.\ NRF-2019R1F1A1062477).

\appendix

\section{The QUMOND effective density of the Miyamoto-Nagai potential under a constant external field} \label{sec:math}

The QUMOND potential for a mass distribution under a constant external field is determined by the effective density of Equation~(\ref{eq:rhohat}) through the usual Poisson equation. Here we describe an analytic expression of the effective density for the MN model (Equation~(\ref{eq:PhiN})). From Equation~(\ref{eq:rhohat}) we have
\begin{equation}
    4\pi\hat{\rho}=\nu\left(|\mathbf{\nabla}\Phi_{\rm N}-e_\text{Ny}\hat{y}-e_\text{Nz}\hat{z}|\right) \mathbf{\nabla}\cdot \mathbf{\nabla}\Phi_{\rm N} +\mathbf{\nabla}\nu\left(|\mathbf{\nabla}\Phi_{\rm N}-e_\text{Ny}\hat{y}-e_\text{Nz}\hat{z}|\right) \cdot (\mathbf{\nabla}\Phi_{\rm N}-e_\text{Ny}\hat{y}-e_\text{Nz}\hat{z})
    \label{eq:4pirho}
\end{equation}
Working out the divergence and the scalar product in Equation~(\ref{eq:4pirho}) we have
\begin{equation}
    4\pi\hat{\rho}  =  4\pi \nu(u)\rho  
      + \frac{\partial\nu(u)}{\partial x} \frac{x}{(R^2+h^2)^{3/2}} + \frac{\partial\nu(u)}{\partial y} \left(\frac{y}{(R^2+h^2)^{3/2}} - e_\text{Ny}\right) + \frac{\partial\nu(u)}{\partial z} \left(\frac{z h} {\sqrt{z^2+b^2}(R^2+h^2)^{3/2}} - e_\text{Nz}\right),
    \label{eq:4pirho1}
\end{equation}
where all lengths are in units of $r_{\rm M}=\sqrt{M}$, $\rho$ is the MN mass density (Equation~(\ref{eq:rho})) normalized to have a total mass of unity, and we have defined $h=h(z)\equiv a+\sqrt{z^2+b^2}$ and
\begin{equation}
    u\equiv|\mathbf{\nabla}\Phi|=\left[ \frac{x^2}{(R^2+h^2)^3} + \left(\frac{y}{(R^2+h^2)^{3/2}} - e_\text{Ny} \right)^2 + \left(\frac{z h}{\sqrt{z^2+b^2}(R^2+h^2)^{3/2}} - e_\text{Nz} \right)^2 \right]^{1/2}.
    \label{eq:u}
\end{equation}
In Equation~(\ref{eq:4pirho1}) the partial derivatives are given as follows:
\begin{equation}
\begin{array}{rl}
      \frac{\partial\nu(u)}{\partial x}  = & \frac{\nu(u)\hat{\nu}(u)}{u^2} \left[ \frac{x}{(R^2+h^2)^3} - \frac{3 x^3}{(R^2+h^2)^4} - \frac{3xy}{(R^2+h^2)^{5/2}} \left(\frac{y}{(R^2+h^2)^{3/2}} - e_\text{Ny}\right) \right. \\
     & \hspace{4em}\left. -\frac{3 x z h}{\sqrt{z^2+b^2}(R^2+h^2)^{5/2}} \left(\frac{z h}{\sqrt{z^2+b^2}(R^2+h^2)^{3/2}} - e_\text{Nz}\right) \right],
\end{array}
    \label{eq:nu_x}
\end{equation} 
\begin{equation}
\begin{array}{rl}
  \frac{\partial\nu(u)}{\partial y}  = & \frac{\nu(u)\hat{\nu}(u)}{u^2} \left[ -\frac{3x^2y}{(R^2+h^2)^4} + \left(\frac{y}{(R^2+h^2)^{3/2}} - e_\text{Ny}\right) \left(\frac{1}{(R^2+h^2)^{3/2}} -  \frac{3y^2}{(R^2+h^2)^{5/2}} \right) \right. \\
     & \hspace{4em}\left. -\frac{3 y z h}{\sqrt{z^2+b^2}(R^2+h^2)^{5/2}} \left(\frac{z h}{\sqrt{z^2+b^2}(R^2+h^2)^{3/2}} - e_\text{Nz}\right) \right],
\end{array}
    \label{eq:nu_y}
\end{equation} 
and
\begin{equation}
\begin{array}{rl}
  \frac{\partial\nu(u)}{\partial z}  = & \frac{\nu(u)\hat{\nu}(u)}{u^2} \left[ -\frac{3x^2 z h}{\sqrt{z^2+b^2}(R^2+h^2)^4} - \frac{3y z h}{\sqrt{z^2+b^2}(R^2+h^2)^{5/2}}  \left(\frac{y}{(R^2+h^2)^{3/2}} - e_\text{Ny}\right) \right. \\
     & \hspace{4em}\left. + \left(\frac{z h}{\sqrt{z^2+b^2}(R^2+h^2)^{3/2}} - e_\text{Nz}\right) \left\{ \frac{h}{\sqrt{z^2+b^2}(R^2+h^2)^{3/2}} + \frac{z^2}{(z^2+b^2)(R^2+h^2)^{3/2}} \left( 1 - \frac{h}{\sqrt{z^2+b^2}} - \frac{3h^2}{R^2+h^2} \right) \right\} \right],
\end{array}
    \label{eq:nu_z}
\end{equation} 
where $\hat{\nu}(u) = d\ln\nu(u)/d\ln u$.
\end{document}